\documentclass[journal, twocolumn]{IEEEtran}
\usepackage{algorithm}
\usepackage{array}

\usepackage{subfigure}
\usepackage{stfloats}
\usepackage{url}
\usepackage{verbatim}
\usepackage{hyperref}

\def\BibTeX{{\rm B\kern-.05em{\sc i\kern-.025em b}\kern-.08em
		T\kern-.1667em\lower.7ex\hbox{E}\kern-.125emX}}

\usepackage{cite}
\usepackage{amsmath,amssymb,amsfonts}
\usepackage{algorithmic}
\usepackage{graphicx}
\usepackage{textcomp}
\usepackage{booktabs}
\usepackage{cuted}
\usepackage{color}
\usepackage{setspace}

\begin{document}

 This manuscript has been submitted to IEEE Antennas and Wireless Propagation Letters.

	\thispagestyle{empty}

\clearpage

\title{A Beam-Steering Reflectarray Antenna with Arbitrary Linear-Polarization Reconfiguration}

\author{Changhao Liu, \IEEEmembership{Graduate Student Member, IEEE}, Songlin Zhou, Fan Yang, \IEEEmembership{Fellow, IEEE}, Shenheng Xu, \IEEEmembership{Member, IEEE} and Maokun Li, \IEEEmembership{Fellow, IEEE}
	\thanks{This work is supported by the National Key Research and Development Program of China under Grant No. 2023YFB3811501. \textit{(Corresponding author: Fan Yang.)}}
	\thanks{The authors are with the Department of Electronic Engineering, State Key Laboratory of Space Network and Communications, Tsinghua University, Beijing 100084, China (e-mail: fan\_yang@tsinghua.edu.cn).}
}

\maketitle

\setcounter{page}{1}

\begin{abstract}
	This work presents a beam-steering reflectarray antenna that achieves arbitrary linear polarization (LP) reconfiguration. This antenna employs a dual-circular polarization (CP) reconfigurable reflectarray and an LP feed horn to generate an LP beam. The incident LP wave is decomposed into two CP components, whose reflection phases are independently adjusted. By re-combining these reflected CP beams with different phase constants, arbitrary LP states can be synthesized. Experimental validation is performed using a 16$\times$16 1-bit reconfigurable reflectarray operating at Ku band, demonstrating reconfigurable LP states of LP(0$^\circ$), LP(45$^\circ$), LP(90$^\circ$) and LP(135$^\circ$), as well as dynamic beam scanning functionality from 0$^\circ$ to 60$^\circ$. This polarization-reconfigurable, beam-steering reflectarray has potential for applications in satellite and mobile communications, where both beam patterns and polarization alignment are crucial.
\end{abstract}


\section{Introduction}
\IEEEPARstart{I}{n} wireless communications, the demand for signal enhancement necessitates the utilization of high-performance antennas. In particular, the alignment of both high-gain beam patterns and polarizations between transmitters (Tx) and receivers (Rx) are vital considerations in antenna design.

Beam-scanning antennas possess the ability to dynamically align beams, which is crucial for point-to-point communications. As a new paradigm of large-aperture antennas, spatial-fed planar reflectarray antennas can generate high-gain beams with significant advantages \cite{reflectarray}. Reconfigurable reflectarray antennas (RRAs) employ tunable elements that enable real-time manipulation of the reflected phase on each element, consequently facilitating dynamic steering of beam patterns \cite{RRAreview1}. The investigation of digital RRAs with 1-bit phase resolution has received significant attention \cite{rra1,rra2,rra3,rra5}. Recent years have witnessed a surge of interest in advanced RRAs with multi-functions \cite{review1}. For example, a dual-polarization-multiplexed RRA is proposed recently, which is capable of independently manipulating the beam patterns for two orthogonal linear polarizations (LPs) \cite{rradualpol}.

Furthermore, numerous practical communication systems rely on LPs. However, as the terminal moves or rotates, the received LP also experiences rotation. Ensuring optimal transmission efficiency necessitates the alignment of polarizations between Tx and Rx. Traditional methods involve mechanical rotation of either the Tx or Rx to achieve polarization alignment, but these approaches suffer from the slow polarization switching speed. Real-time polarization alignment requires dynamic polarization tuning techniques in antenna design. Recently, polarization reconfigurable surfaces can generate tunable polarizations. By integrating lumped switches on the elements, the scattered polarization states can be dynamically controlled, as reported in \cite{polLPLP,polLPCP,polCPCP,polarb2}.

To address the dual requirements of beam and polarization alignment in wireless communications, it is essential to develop an antenna that possesses both beam-steering and polarization-reconfigurable capabilities.

Phase constant, or reference phase, is important in reflectarray designs \cite{reflectarray}. Advanced functions can be achieved by tuning phase constants, like gain improvement \cite{gain}, bandwidth enhancement \cite{bandwidth}, and sidelobe suppression \cite{sidelobe}. Phase constant can also be applied in polarization-reconfigurable beam-steering reflectarrays. Dual-LP 1-bit RRAs are employed to synthesize the desired polarization at the target beam direction, whereby tuning the phase constants of the two LP channels can change the polarization states \cite{polbeam2,polbeam3}. However, these designs only generate HLP, VLP, LHCP and RHCP beams, but do not allow for the generation of arbitrary LPs.

In this paper, we propose an RRA that enables arbitrary LP reconfiguration. By leveraging a dual-CP 1-bit RRA, arbitrary LPs can be synthesized by manipulating the relative phase of the dual-CP directional waves. As a demonstration of concept, a 16$\times$16 dual-CP 1-bit RRA operating at 16.8 GHz is employed to generate scanning beams corresponding to LP(0$^\circ$), LP(45$^\circ$), LP(90$^\circ$) and LP(135$^\circ$), respectively. For each polarization state, the gains and the beam scanning patterns are measured, demonstrating the precise tuning of beam directions and arbitrary LPs as desired. This paper is organized as follows. Section II proposes the operation principle of arbitrary LP synthesis. Section III shows the experimental demonstration of the proposed  method. Finally, the conclusion is drawn in Section IV.

\section{Operation Principle}
\subsection{Relations Between the Phase Constant and the Wave Phase}
\label{secA}

To begin with, we establish a lemma: The phase of the reflected EM wave exhibits a near-linear relationship with the phase constant of the 1-bit RRA.

Let us consider a reflectarray composed of $M\times N$ elements, which is fed by a near-field horn. The reflection from the $(m, n)$th element possesses a compensation phase denoted as $\varphi_{mn}$ and a magnitude represented by $E_{mn}$. Summing all the reflected waves, the total EM wave at the direction $\vec{k}=\left(\theta_0, \phi_0\right)$ can be expressed as
\begin{equation}
		E=\frac{e^{-j\vec{k} \cdot \vec{d}}}{|d|}\sum_{m=1}^{M}\sum_{n=1}^{N}E_{mn}e^{-j(\varphi_{f m n}-\varphi_{m n}-\vec{k} \cdot \vec{r}_{mn})}.
\end{equation}
Here, $\varphi_{f m n}$ represents the phase delay from the feed to the $(m, n)$th element, $\vec{r}_{mn}$ denotes the location of the element, and $d$ is the distance between the receiver and the array. 

To achieve maximum gain at $\vec{k}$ direction, the required compensation phase for the $(m, n)$th element should satisfy
\begin{equation}
	\varphi^{req}_{m n}=\varphi_{f m n}-\vec{k} \cdot \vec{r}_{mn}+\Delta\varphi,
\end{equation}
where $\Delta\varphi$ is defined as the phase constant of the reflectarray \cite{reflectarray}. If the reflectarray elements can produce continuous compensation phases, the required phase $\varphi^{req}_{m n}$ can be obtained, with $\varphi_{m n}=\varphi^{req}_{m n}$. Consequently, the total EM wave is
\begin{equation}
	E=e^{j\Delta\varphi}\frac{e^{-j\vec{k} \cdot \vec{d}}}{|d|}\sum_{m=1}^{M}\sum_{n=1}^{N}E_{mn}.
\end{equation}
The magnitude of the total EM wave is not affected by $\Delta\varphi$. However, the variation in the phase constant results in a linear variation in the phase of the EM wave, denoted by $\varphi_w$.

For 1-bit RRAs, only two values of $\varphi_{m n}$ are available. Suppose $\varphi_{m n}$ can be quantized as 0$^\circ$ or 180$^\circ$, and the phase quantization strategy is
\begin{equation}
	\varphi_{m n}=
	\left \{
	\begin{array}{lcl}
		0^\circ & & -90^\circ\leq\varphi^{req}_{m n}<90^\circ \\
		180^\circ & & 90^\circ\leq\varphi^{req}_{m n}<270^\circ
	\end{array}
	\right.
\end{equation} 
We further introduce the phase quantization error
\begin{equation}
	\varphi^{err}_{m n} = \varphi_{m n} - \varphi^{req}_{m n},
\end{equation}
where $\varphi^{err}_{m n}\in[-90^\circ, 90^\circ]$. Hence, the total EM wave is simplified as
\begin{equation}
	E=e^{j\Delta\varphi}\frac{e^{-j\vec{k} \cdot \vec{d}}}{|d|}\sum_{m=1}^{M}\sum_{n=1}^{N}E_{mn}e^{j\varphi^{err}_{m n}}.
\end{equation}
It has been observed that the distribution of the phase quantization errors is pseudorandom within the range of $[-90^\circ, 90^\circ]$ \cite{quant}. According to the law of large numbers, as $M$ and $N$ increase, the phase of $\sum_{m=1}^{M}\sum_{n=1}^{N}E_{mn}e^{j\varphi^{err}_{m n}}$ converges towards 0$^\circ$. Thus, even under the 1-bit phase quantization, the overall phase of the EM wave still exhibits a near-linear variation with the phase constant.

\begin{figure} 
	\centering 
	 \label{simwavephase} 
		\includegraphics[width=0.45\columnwidth]{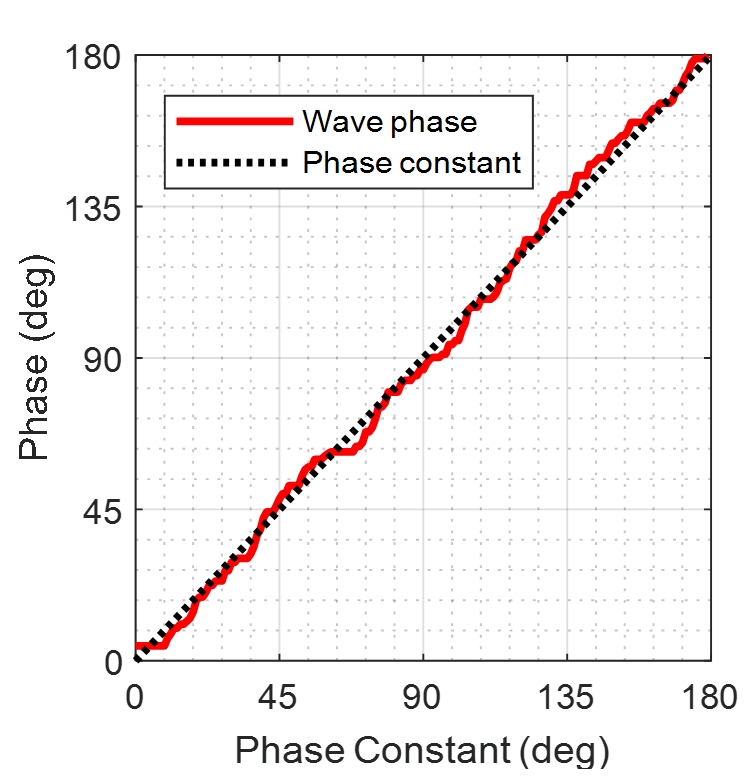} 
	 
	\caption{Simulated Variation of reflection wave phases with respect to the phase constant of a 16$\times$16 1-bit RRA.} 
\end{figure}

The simulation results based on the 16$\times$16 1-bit RRA provide evidence supporting the theory that the wave phase changes near linearly with the phase constant, as shown in Fig. \ref{simwavephase}. However, due to the pseudorandom property of 1-bit phase quantization and the limited size of the array, there is a phase error which is within 6$^\circ$.

In summary, it is found that the phase constant $\Delta\varphi$ serves as a new degree of freedom at the array level to tune the phase of the reflected EM wave $\varphi_w$ from the 1-bit RRA.

\subsection{Arbitrary LP Synthesis Using Phase Constants Based on Dual-CP RRA}

\begin{figure} 
	\centering 
	\subfigure[] { \label{sche} 
		\includegraphics[width=0.65\columnwidth]{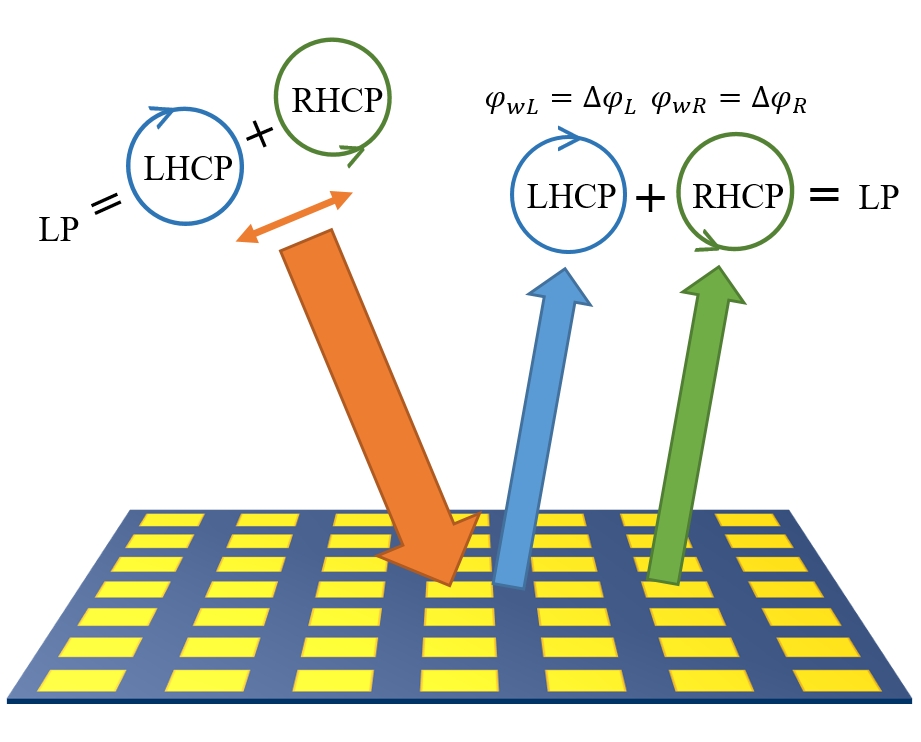} 
	} 
	\subfigure[] { \label{compose} 
		\includegraphics[width=0.65 \columnwidth]{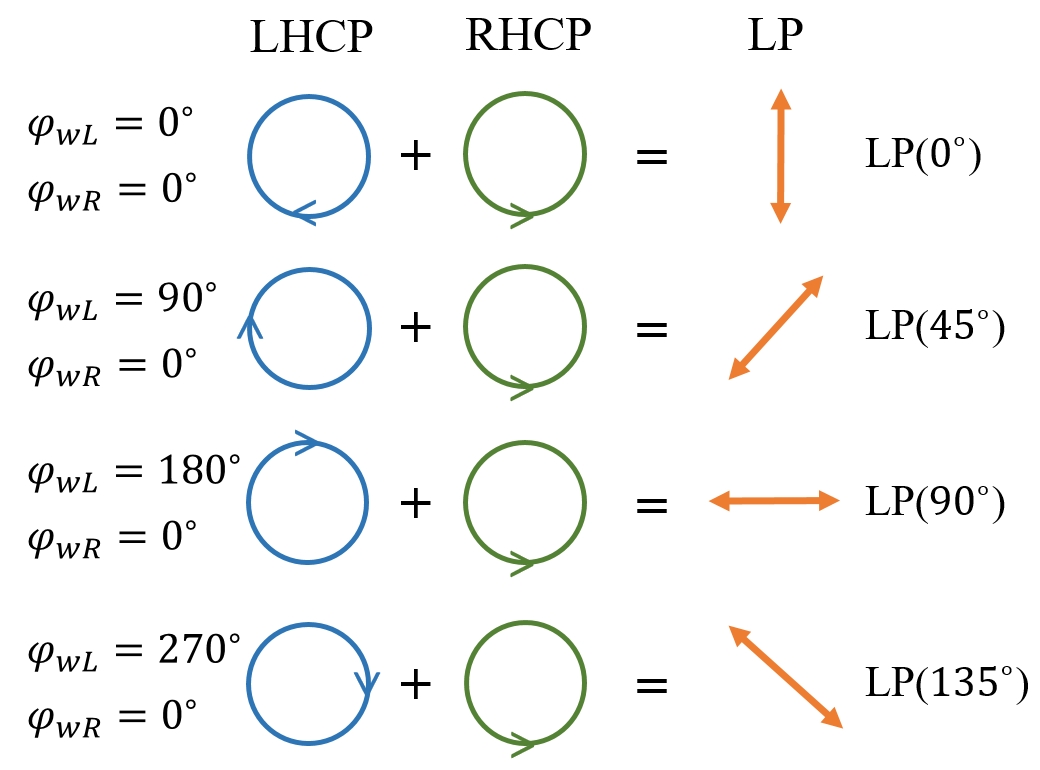} 
	} 
	\caption{Schematic illustrations of arbitrary LP generation utilizing the dual-CP RRA. (a) Schematic depicting the generation of dual-CP scanning beams using the LP feed and the dual-CP RRA. (b) Generation of arbitrary LP states by manipulating the relative phases between dual-CP waves.} 
	\label{sches} 
\end{figure}

Suppose there exists a dual-CP 1-bit RRA capable of steering the LHCP and RHCP beams independently. Any arbitrary LP is composed of LHCP and RHCP components with identical amplitudes, but different phases. When an LP wave illuminates on the dual-CP RRA, the decomposed LHCP and RHCP beams can be manipulated independently, as shown in Fig. \ref{sche}. Then, let us examine a scenario in which the LHCP and RHCP scanning beams are directed towards the same direction. Mathematically, the LHCP and RHCP waves are represented as
\begin{equation}
	\vec{E}_L = A_Le^{j\varphi_{wL}}
	\begin{pmatrix}
		1\\
		-j
	\end{pmatrix}, 	
	\vec{E}_R = A_Re^{j\varphi_{wR}}	
	\begin{pmatrix}
		1\\
		j
	\end{pmatrix},
\end{equation}
where $\varphi_{wL}$ and $\varphi_{wR}$ the phase of LHCP and RHCP waves, respectively. Suppose the amplitudes are equal, denoted as $A_0 = A_L = A_R$. Without loss of generality, let $\varphi_{wR}=0^\circ$. Combining the two CP waves, the synthesized wave is expressed as
\begin{equation}
	\begin{aligned}
		\vec{E} & = \vec{E}_L+\vec{E}_R=A_0
		\begin{pmatrix}
			e^{j\varphi_{wL}}+1 \\
			-je^{j\varphi_{wL}}+j
		\end{pmatrix} \\
		&= 2A_0e^{j\varphi_{wL}/2}
		\begin{pmatrix}
			\cos(\varphi_{wL}/2) \\
			\sin(\varphi_{wL}/2)
		\end{pmatrix},
	\end{aligned}
\end{equation}
which represents an LP wave with polarization along $\varphi_{wL}/2$. This demonstrates that the combination of these two CP waves can generate an LP wave without losing efficiency, and the LP state is determined by $\varphi_{wL}$. Geometrically, Fig. \ref{compose} illustrates the generation of LP(0$^\circ$), LP(45$^\circ$), LP(90$^\circ$) and LP(135$^\circ$) waves by fixing the phase of the RHCP wave and adjusting the phase of the LHCP wave. 

As discussed in Section \ref{secA}, the phase constants of the dual-CP RRA can linearly manipulate the phases of the dual-CP waves. Therefore, by selecting different phase constants, it becomes feasible to generate arbitrary LP states. 

\section{Experimental Demonstration}

\subsection{Prototype Design And Performances}

\begin{figure}[!t]
	\centerline{\includegraphics[width=0.7\columnwidth]{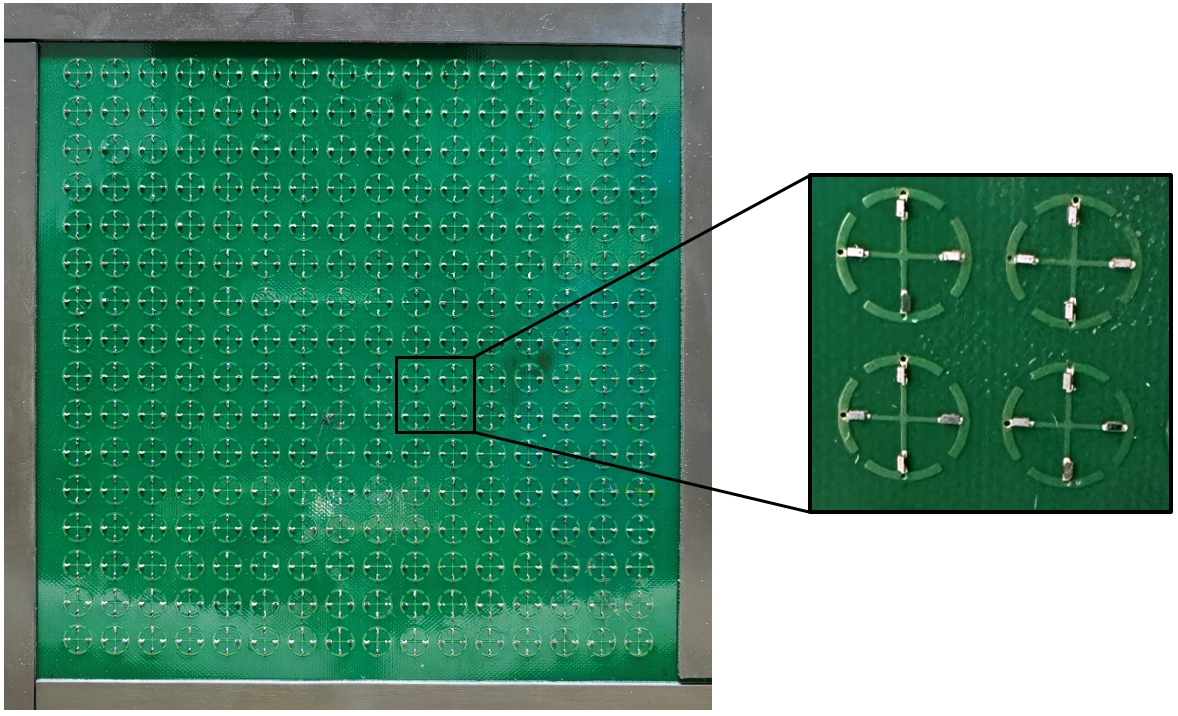}}
	\caption{The 16$\times$16 dual-CP 1-bit RRA prototype.}
	\label{prototype}
\end{figure}

A 16$\times$16-element Ku-band dual-CP 1-bit RRA prototype is designed and fabricated in our previous work \cite{CPRRA}, as shown in Fig. \ref{prototype}. The period of the element is 8 mm, and the aperture size is 12.8 mm $\times$ 12.8 mm. The top substrate is TLX-8 with a thickness of 1.57 mm, while the bottom substrate is FR-4 with a thickness of 1.4 mm for placing biasing lines, and they are bonded together by a prepreg layer. The reflective ground plane is on the bottom side of the top substrate. Each element integrates four PIN diodes, of which three can be controlled independently. This design allows the element to independently respond to incident waves of LHCP and RHCP, with the ability to achieve 1-bit phase tuning. Details of the structure design can be found in \cite{CPRRA}.

\begin{figure} 
	\centering 
	\subfigure[] { \label{measphaseL} 
		\includegraphics[width=0.46\columnwidth]{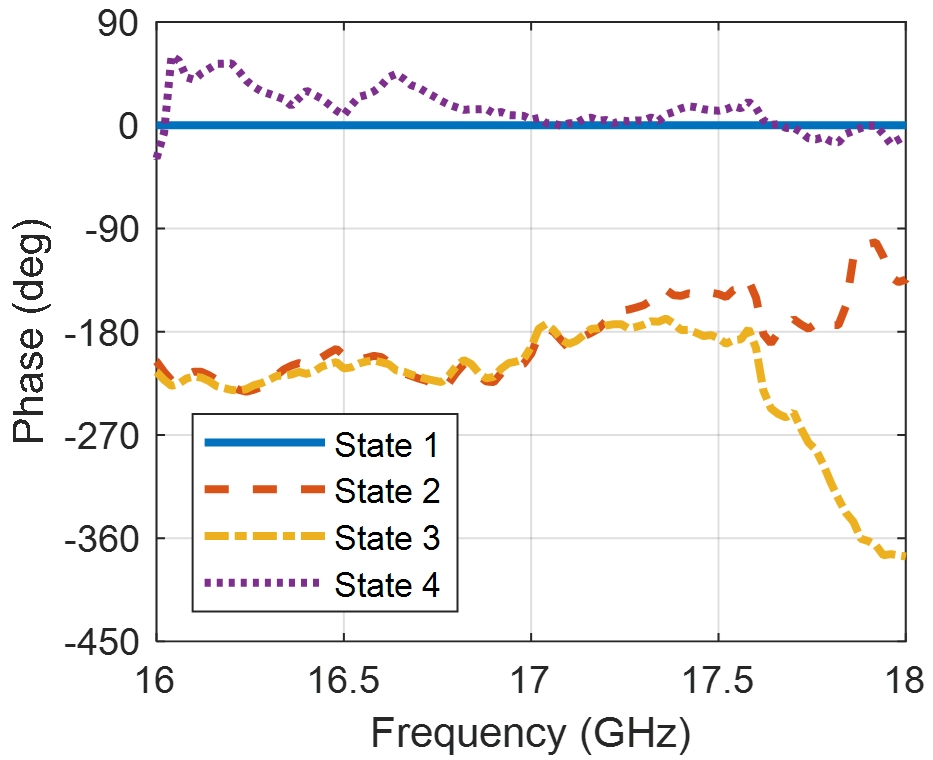} 
	} 
	\subfigure[] { \label{measphaseR} 
		\includegraphics[width=0.46\columnwidth]{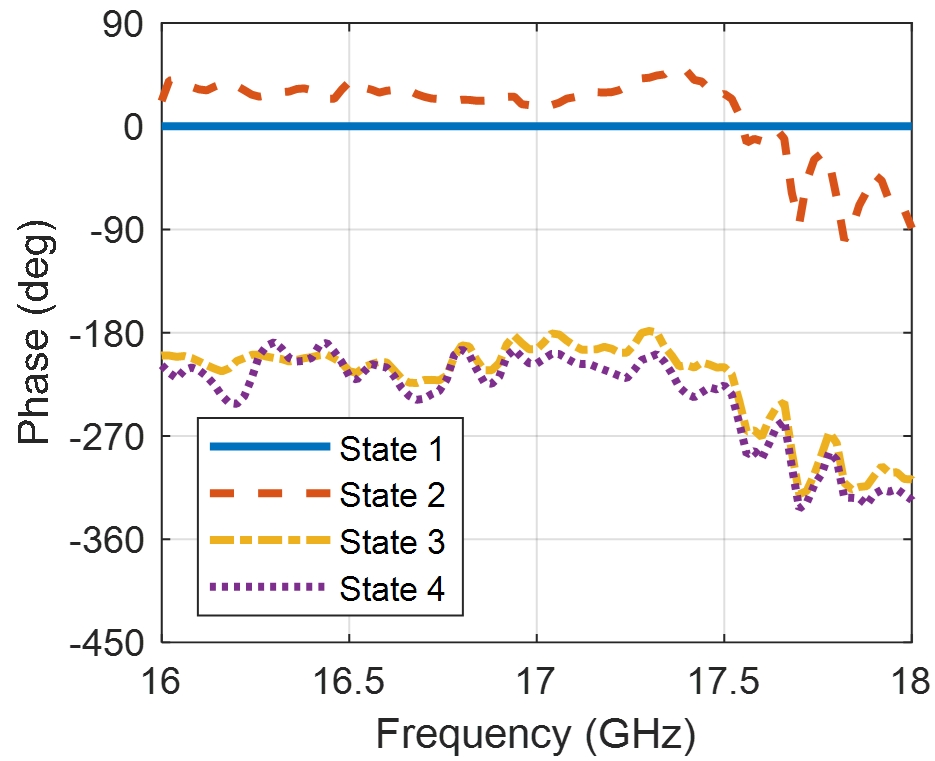} 
	} 
	\caption{Measured reflection phases for LHCP and RHCP waves of the element. The phases of State 1 are normalized to 0$^\circ$. (a) Reflection phases for LHCP wave. (b) Reflection phases for RHCP wave.} 
	\label{measelement} 
\end{figure}

The measured results of the element performance are shown in Fig. \ref{measelement}. The dual-CP 1-bit element offers four distinct states with independent response to two CP waves, denoted as State 1: LHCP-0, RHCP-0; State 2: LHCP-1, RHCP-0; State 3: LHCP-1, RHCP-1; State 4: LHCP-0, RHCP-1. It operates at a central frequency of 16.8 GHz. For LHCP incident and reflected waves, when solely adjusting the LHCP states (from State 1 to State 2/3), the reflection phase differences consistently jump to 180$^\circ$, whereas only altering the RHCP states (from State 1 to State 4 or from State 2 to State 3) results in reflection phase variance close to 0$^\circ$. This behavior is similar for RHCP incident waves as well. These results demonstrate that the dual-CP element can independently manipulate the LHCP and RHCP waves with 180$^\circ$ phase shift, while simultaneously maintaining minimal mutual coupling between the two polarizations.

Literature \cite{CPRRA} has effectively demonstrated the independent beam-scanning capabilities for both LHCP and RHCP waves. In this work, we leverage this prototype to show its the beam-scanning ability with the arbitrary LP synthesis function.

\subsection{Verification of the Phase Constant Effect}

Firstly, the linear relationship between the wave phase and the phase constant is experimentally verified. The dual-CP RRA is fed by a dual-CP horn antenna. This RRA produces dual-CP pencil-beams directed toward the broadside. A second dual-CP horn, positioned in the far-field of the RRA along the broadside direction, is employed to measure the reflection phases of the 0$^\circ$ beam. The phase constants are varied from 0$^\circ$ to 180$^\circ$ for both LHCP and RHCP cases.

\begin{figure} 
	\centering 
	\subfigure[] { \label{measLphase} 
		\includegraphics[width=0.46\columnwidth]{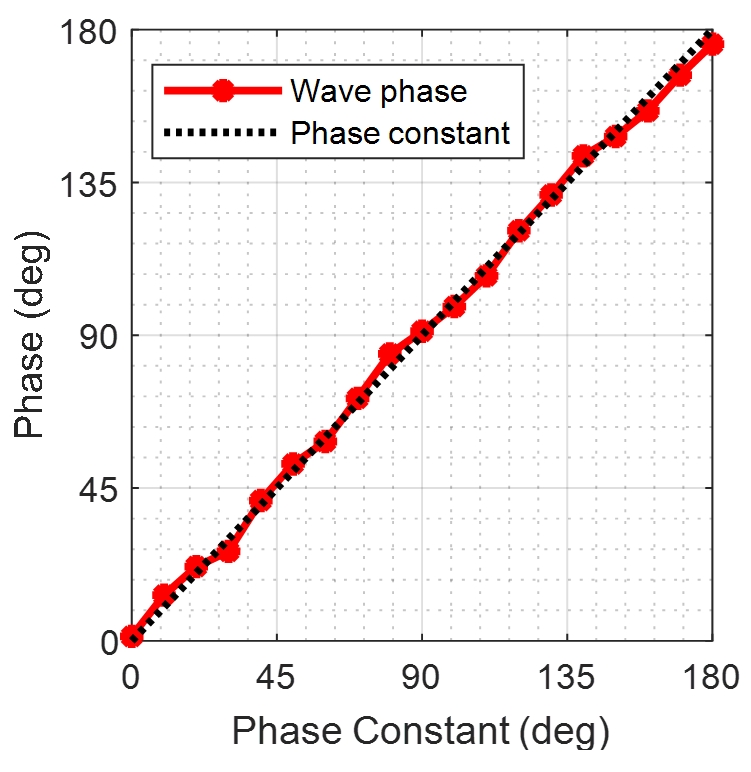} 
	} 
	\subfigure[] { \label{measRphase} 
		\includegraphics[width=0.46\columnwidth]{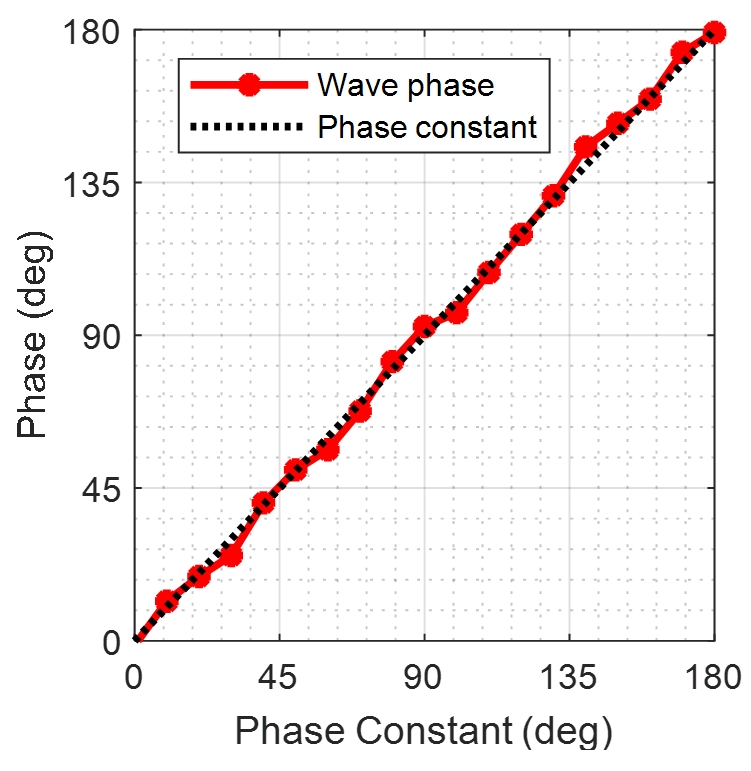} 
	} 
	\caption{Measured phases of LHCP and RHCP waves at 16.8 GHz with varying phase constants. (a) Phase of LHCP wave. (b) Phase of RHCP wave.} 
	\label{measwavephase} 
\end{figure}

The received phases are measured at 16.8 GHz. Fig. \ref{measwavephase}(a) and (b) show the relationship between the wave phases and the phase constants for LHCP and RHCP waves, respectively. The root mean squared errors (RMSEs) between the wave phases and phase constants are found to be 2.4$^\circ$ for the LHCP wave and 2.5$^\circ$ for the RHCP wave, providing evidence of an excellent linear relationship between the wave phase and phase constants.

\subsection{Polarization-Reconfigurable Beamforming Measurement}

\begin{figure}[!t]
	\centerline{\includegraphics[width=0.65\columnwidth]{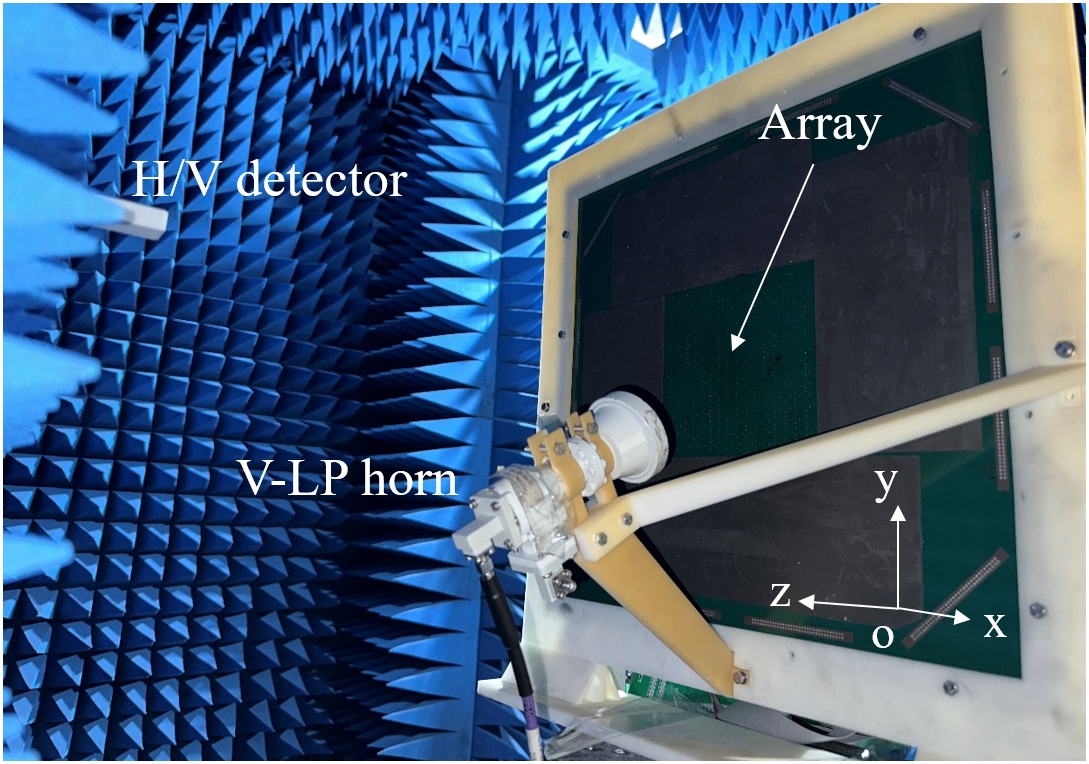}}
	\caption{Measurement setup for polarization-reconfigurable beam steering.}
	\label{beammeas}
\end{figure}

The beamforming and beam-scanning measurements are conducted in a near-field microwave anechoic chamber. As shown in Fig. \ref{beammeas}, the dual-CP RRA is fed by a VLP horn at an oblique incident angle of 30$^\circ$, with an F/D ratio of 1. The detector moves in xoy plane to measure the field in two orthogonal LPs and synthesize the final detected polarization. To obtain the broadside beam patterns for LPs, both RHCP and LHCP pencil-beams are directed towards 0$^\circ$. The phase constant for RHCP wave is fixed at 0$^\circ$, while the phase constants for LHCP wave vary from 0$^\circ$, 90$^\circ$, 180$^\circ$ to 270$^\circ$, in order to generate LP(0$^\circ$), LP(45$^\circ$), LP(90$^\circ$) and LP(135$^\circ$) beams, respectively.

\begin{figure} 
	\centering 
\subfigure[] { \label{broadside90xoz} 
		\includegraphics[width=0.46\columnwidth]{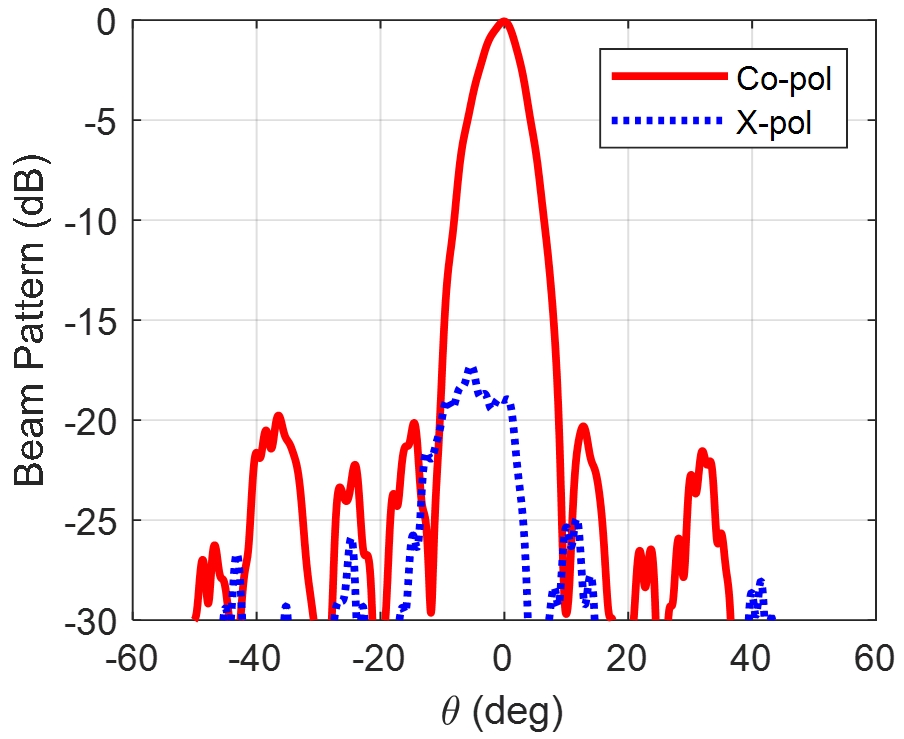} 
	} 
	\subfigure[] { \label{broadside45xoz} 
	\includegraphics[width=0.46\columnwidth]{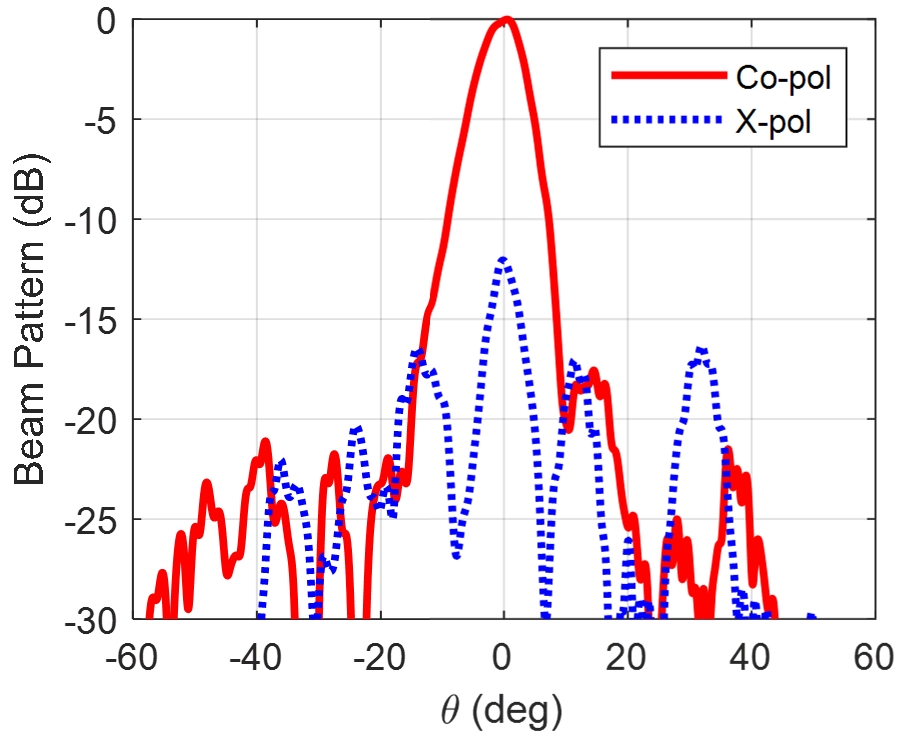} 
	} 
\subfigure[] { \label{broadside0xoz} 
	\includegraphics[width=0.46\columnwidth]{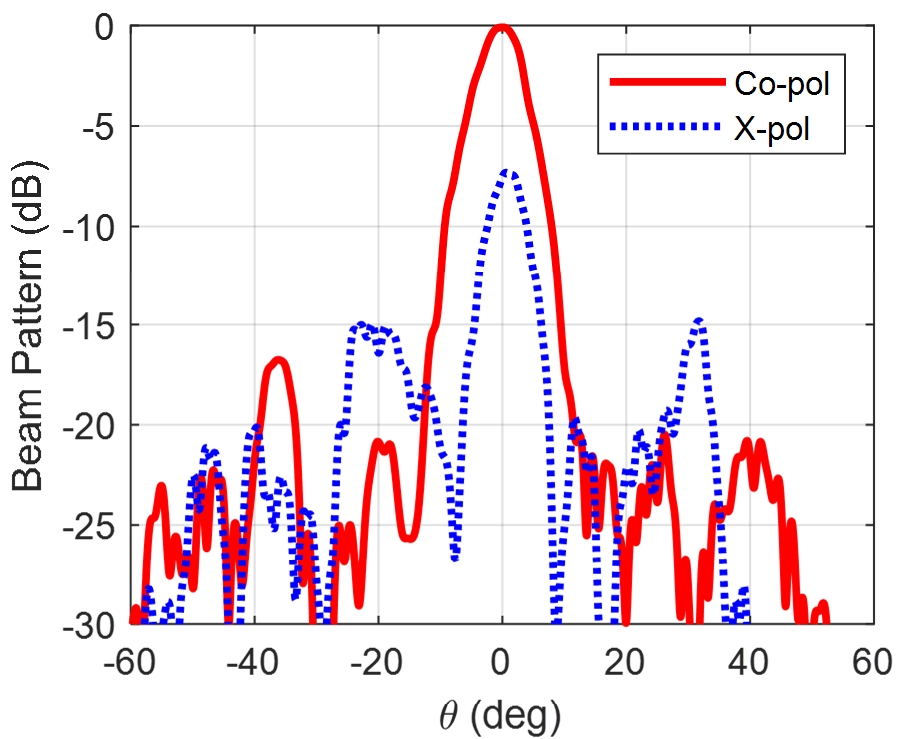} 
} 
	\subfigure[] { \label{broadside135xoz} 
	\includegraphics[width=0.46\columnwidth]{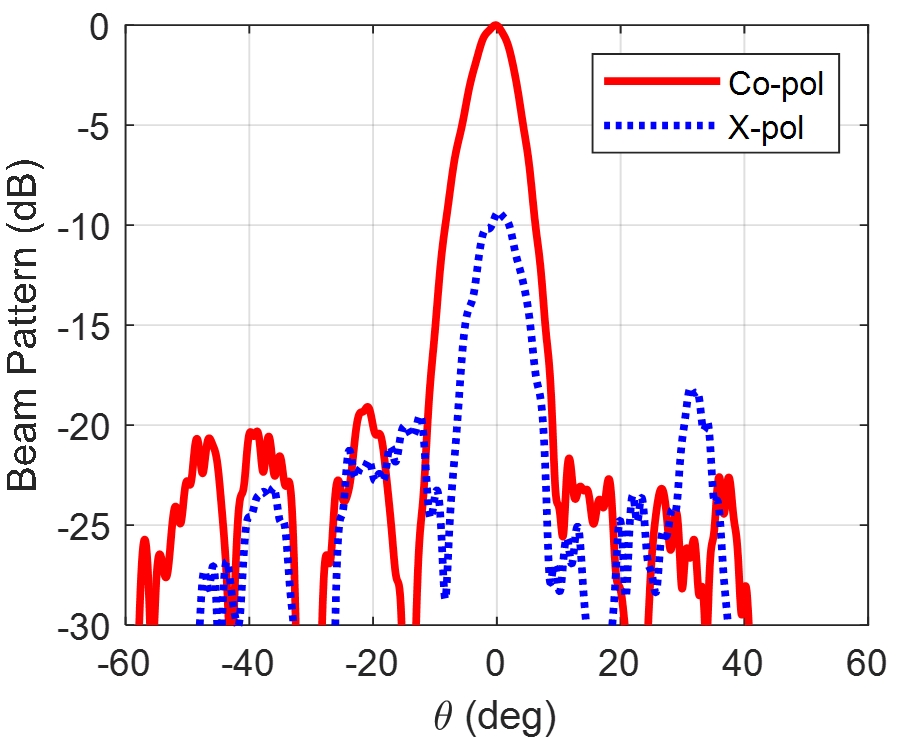} 
} 
	\caption{Measured broadside beams at 16.8 GHz. (a) LP(0$^\circ$). (b) LP(45$^\circ$). (c) LP(90$^\circ$). (d) LP(135$^\circ$).} 
	\label{broadside} 
\end{figure}

The measured broadside beam patterns at 16.8 GHz for various LPs are shown in Fig. \ref{broadside}. Clear and well-defined pencil-beams are observed across different polarizations in xoz plane, demonstrating the ability to achieve arbitrary LP configurations as desired. The cross-polarization level remains below -7.3 dB at a beam angle of 0$^\circ$ for all LPs, indicating acceptable polarization purity. 

\begin{figure}[!t]
	\centerline{\includegraphics[width=0.6\columnwidth]{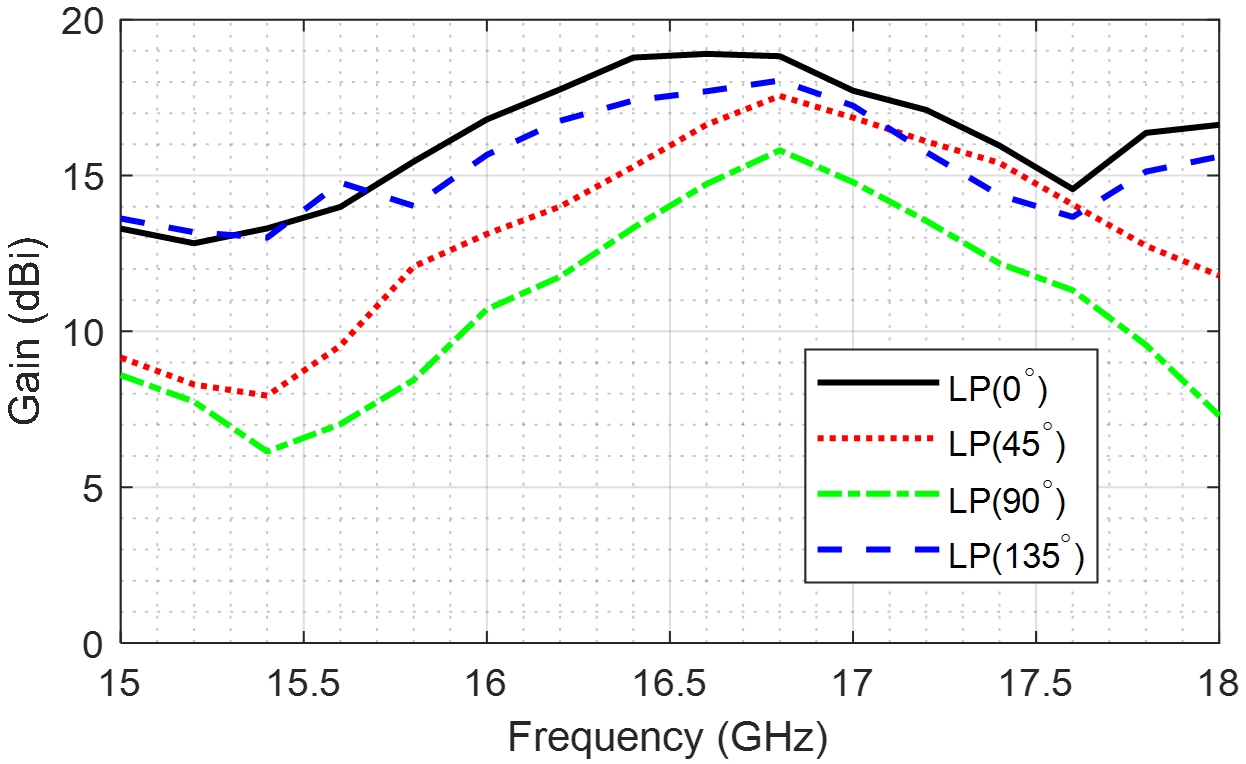}}
	\caption{Measured gains of broadside beams for LP(0$^\circ$), LP(45$^\circ$), LP(90$^\circ$) and LP(135$^\circ$). }
	\label{gain}
\end{figure}

The gains of the LP(0$^\circ$), LP(45$^\circ$), LP(90$^\circ$) and LP(135$^\circ$) broadside beams are measured from 15 GHz to 18 GHz, as shown in Fig. \ref{gain}. It is observed that the gains peak at 16.8 GHz for all LP states. At this frequency, the LP(0$^\circ$) beam achieves a maximum gain of 18.8 dBi. The LP(90$^\circ$) beam exhibits the minimum gain of 15.8 dBi. The range of gain variation is 3 dB, primarily influenced by the 1-bit phase quantization and the non-ideal variation of phase and amplitudes of two CP beams. Furthermore, the LP(0$^\circ$) beam has the maximum 3-dB gain bandwidth of 9.2\%, while the LP(0$^\circ$) beam demonstrates a minimum 3-dB gain bandwidth of 5.8\%, which aligns with the bandwidth of the dual-CP RRA previously described \cite{CPRRA}.

The cross polarization levels, sidelobes and non-uniform gains are mainly caused by the fluctuation of gains between RHCP and LHCP waves and the wave phase errors. In the future, increasing the size of the RRA and improving the phase quantization resolution can effectively address these issues.

\subsection{Beam Scanning Measurement}

\begin{figure} 
	\centering 
	\subfigure[] { \label{scan90xoz} 
		\includegraphics[width=0.46\columnwidth]{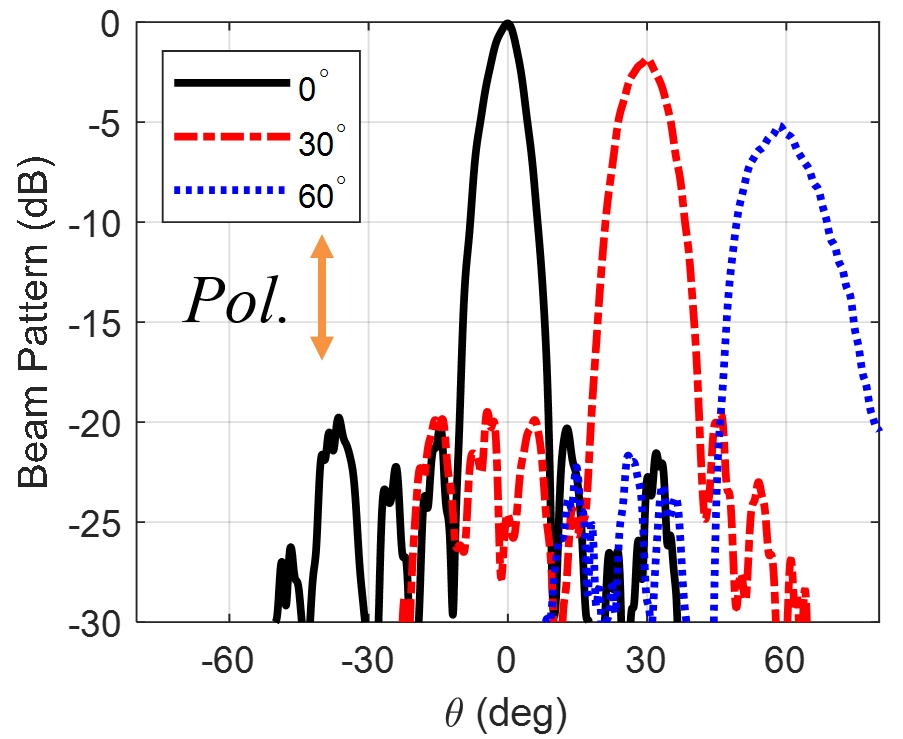} 
	} 
	\subfigure[] { \label{scan45xoz} 
		\includegraphics[width=0.46\columnwidth]{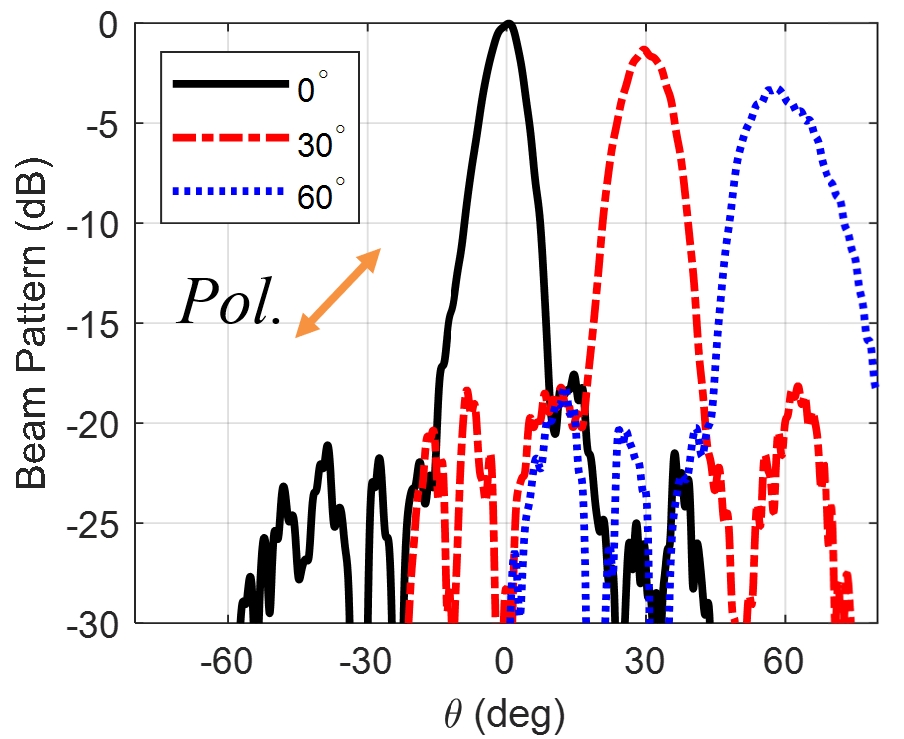} 
	}
	\subfigure[] { \label{scan0xoz} 
		\includegraphics[width=0.46\columnwidth]{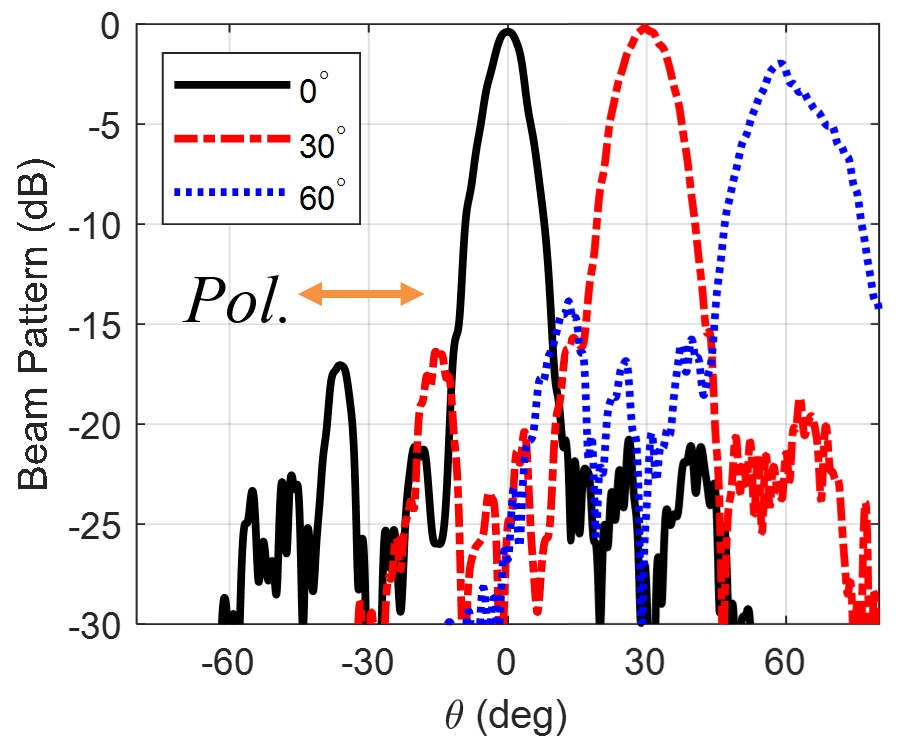} 
	} 
	\subfigure[] { \label{scan135xoz} 
		\includegraphics[width=0.46\columnwidth]{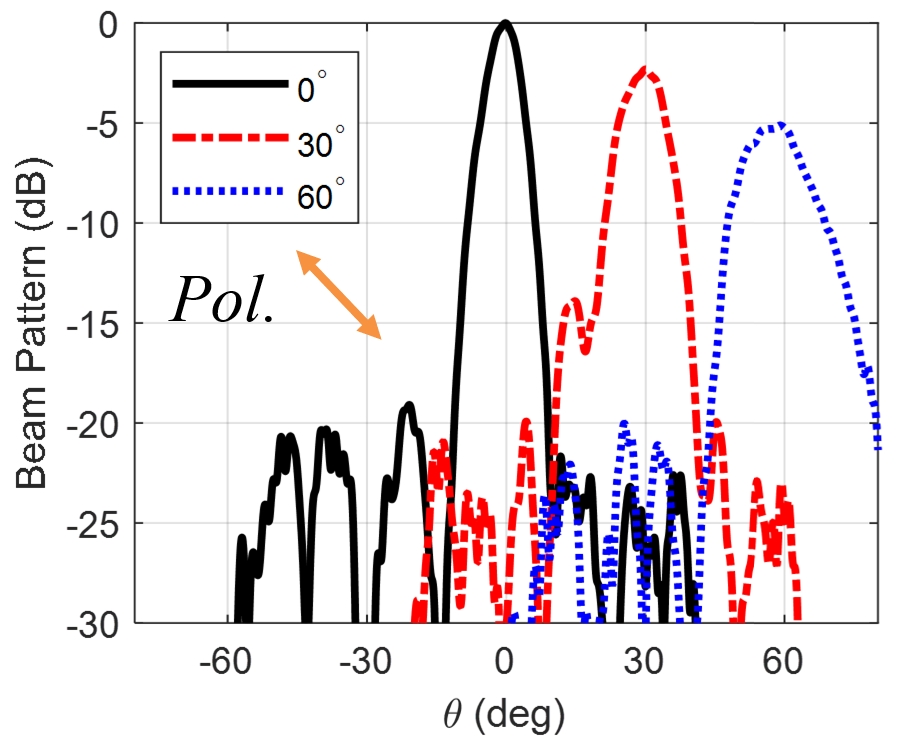} 
	} 
	\caption{Measured scanning beams of (a) LP(0$^\circ$), (b) LP(45$^\circ$), (c) LP(90$^\circ$), (d) LP(135$^\circ$).} 
	\label{scanxoz} 
\end{figure}

The beam scanning patterns for the LP(0$^\circ$), LP(45$^\circ$), LP(90$^\circ$) and LP(135$^\circ$) are measured by electrically changing the coding patterns of the RRA. Fig. \ref{scanxoz} illustrates the beam scanning patterns in the xoz plane, showing that all LPs can support scanning up to 60$^\circ$, with a maximum scan loss of 5.18 dB for LP(0$^\circ$). The sidelobe levels are below -11 dB. Additionally, there is no grating lobe at any scanning angle, which is suitable for large angle beam scanning at arbitrary LPs. Similarly, the polarization-reconfigurable scanning beams can also support up to 60$^\circ$ in yoz plane.

\section{Conclusion}

This paper presents a 1-bit RRA that offers both arbitrary LP reconfiguration and beam scanning capabilities. The study first explores the linear relationship between wave phase and phase constant, where the phase constant serves as a novel degree of freedom for manipulating EM wave at the array level. Subsequently, LPs are synthesized by manipulating the LHCP and RHCP directional beams based on the dual-CP RRA. By adjusting the phase constants of the LHCP and RHCP waves, the arbitrary LP states can be generated as desired. Experimental validations are conducted to verify the proposed method, wherein a dual-CP 1-bit RRA prototype successfully synthesizes reconfigurable LP(0$^\circ$), LP(45$^\circ$), LP(90$^\circ$), and LP(135$^\circ$) pencil-beams. Furthermore, beam scanning patterns are measured for the four polarization states, revealing a 60$^\circ$ beam scanning range. The proposed LP-reconfigurable RRA exhibits great potential for wireless communications, particularly in scenarios requiring the alignment for both high-gain beam and polarizations simultaneously.


\begin{thebibliography}{00}
	\bibitem{reflectarray}  P. Nayeri, F. Yang, and A. Z. Elsherbeni, \emph{Reflectarray Antennas: Theory, Designs and Applications.} New York, NY, USA: Wiley, 2018.
	
	\bibitem{RRAreview1} S. V. Hum and J. Perruisseau-Carrier, ``Reconfigurable reflectarrays and array lenses for dynamic antenna beam control: A review,'' \emph{IEEE Trans. Antennas Propag.}, vol. 62, no. 1, pp. 183–198, Jan. 2014.
	
	\bibitem{rra1} H. Kamoda, T. Iwasaki, J. Tsumochi, T. Kuki and O. Hashimoto, ``60-GHz electronically reconfigurable large reflectarray using single-bit phase shifters,''  \emph{IEEE Trans. Antennas Propag.}, vol. 59, no. 7, pp. 2524-2531, July 2011.
	\bibitem{rra2} H. Yang~\emph{et al.}, ``A 1-bit $10 \times 10$ reconfigurable reflectarray antenna: design, optimization, and experiment,'' \emph{IEEE Trans. Antennas Propag.}, vol. 64, no. 6, pp. 2246-2254, June 2016.
	\bibitem{rra3} J. Han, L. Li, G. Liu, Z. Wu and Y. Shi, ``A wideband 1 bit $12\times12$ reconfigurable beam-scanning reflectarray: Design fabrication and measurement,'' \emph{IEEE Antennas Wireless Propag. Lett.}, vol. 18, no. 6, pp. 1268-1272, Jun. 2019.
	\bibitem{rra5} X. Pan, F. Yang, S. Xu and M. Li, ``A 10240-element reconfigurable reflectarray with fast steerable monopulse patterns,'' \emph{IEEE Trans. Antennas Propag.}, vol. 69, no. 1, pp. 173-181, Jan. 2021.
	
	
	\bibitem{review1} C. Liu, F. Yang, S. Xu, and M. Li, ``Reconfigurable metasurface: A systematic categorization and recent advances,'' \emph{Electromagn. Sci.}, vol. 1, no. 4, pp. 1-23, Dec. 2023.
	
	\bibitem{rradualpol} N. Zhang~\emph{et al.}, ``A dual-polarized reconfigurable reflectarray antenna based on dual-channel programmable metasurface,'' \emph{IEEE Trans. Antennas Propag.}, vol. 70, no. 9, pp. 7403-7412, Sept. 2022.
	
	\bibitem{polLPLP} H. Yang~\emph{et al.}, ``A programmable metasurface with dynamic polarization, scattering and focusing control,'' \emph{Sci. Rep.}, vol. 6, no. 35692, pp. 1–11, Oct. 2016.
	\bibitem{polLPCP} H. Yu~\emph{et al.}, ``Design of a wideband and reconfigurable polarization converter using a manipulable metasurface,'' \emph{Opt. Mater. Exp.}, vol. 8, no. 11, pp. 3373-3381, 2018.
	\bibitem{polCPCP} X. Ma~\emph{et al.}, ``An active metamaterial for polarization manipulating,'' \emph{Adv. Opt. Mater.}, vol. 2, no. 10, pp. 945–949, Oct. 2014.
	
	\bibitem{polarb2} Q. Hu~\emph{et al.}, ``On-demand dynamic polarization meta-transformer,'' \emph{Laser. Photonics. Rev.}, vol. 17, no. 1, Jan. 2023.
	
	
	\bibitem{gain} W. Li~\emph{et al.}, ``A reflectarray with low profile based on the method of choosing the certain phase constant,'' \emph{IEEE Antennas Wireless Propag. Lett.}, in press.
	
	
	\bibitem{bandwidth} Y. Mao, S. Xu, F. Yang, and A. Z. Elsherbeni, ``A novel phase synthesis approach for wideband reflectarray design,'' \emph{IEEE Trans. Antennas Propag.}, vol. 63, no. 9, pp. 4189–4193, Sep. 2015.
	
	\bibitem{sidelobe} X. Chen~\emph{et al.}, ``A 265-GHz CMOS reflectarray with 98$\times$98 elements for 1$^\circ$-wide beam forming and high-angular-resolution radar imaging,'' \emph{IEEE J. Solid-St. Circ.}, vol. 59, no. 11, pp. 3655-3669, Nov. 2024.
	
	\bibitem{polbeam2} J. Hu, P. -L. Chi and T. Yang, ``Novel 1-bit beam-scanning reflectarray with switchable linear, left-handed, or right-handed circular polarization,'' \emph{IEEE Trans. Antennas Propag.}, vol. 71, no. 2, pp. 1548-1556, Feb. 2023.
	
	\bibitem{polbeam3} H. Yu et al., ``Quad-polarization reconfigurable reflectarray with independent beam scanning and polarization switching capabilities,'' \emph{IEEE Trans. Antennas Propag.}, vol. 71, no. 9, pp. 7285–7298, Sep. 2023.

	\bibitem{quant} H. Yang~\emph{et al.}, ``A study of phase quantization effects for reconfigurable reflectarray antennas,'' \emph{IEEE Antennas Wireless Propag. Lett.}, vol. 16, no. 5, pp. 302-305, Mar. 2017.
	
	\bibitem{CPRRA} S. Zhou, F. Yang, S. Xu, M. Li, ``A dual-circularly polarized reconfigurable reflectarray antenna with independent beam scanning capability,'' \emph{IEEE Trans. Antennas Propag.}, vol. 72, no. 9, pp. 7100-7109, Sep. 2024.
	
\end{thebibliography}
\end{document}